\documentstyle[12pt,aaspp4]{article}

\lefthead{Bekki, Couch, \& Drinkwater}
\righthead{Galaxy threshing}

\begin{document}
\title{Galaxy threshing and the formation of ultra-compact dwarf galaxies}

\author{Kenji Bekki and Warrick J. Couch} 
\affil{
School of Physics, University of New South Wales, Sydney 2052, Australia}

\and

\author{Michael  J. Drinkwater}
\affil{
School of Physics, University of Melbourne,  Victoria 3010, Australia}

\begin{abstract}
Recent spectroscopic and morphological observational studies of  galaxies
around NGC~1399 in the Fornax Cluster (Drinkwater et al. 2000b) 
have discovered several `ultra-compact dwarf'
galaxies with intrinsic sizes of $\sim$ 100 pc and absolute
$B$ band magnitudes ranging from $-13$ to $-11$ mag.
In order to elucidate the origin of these enigmatic objects,
we perform numerical simulations on the dynamical evolution
of  nucleated dwarf galaxies orbiting NGC~1399 and suffering from its
strong tidal gravitational field. Adopting a plausible scaling
relation for dwarf galaxies, we find that the outer stellar components of
a nucleated dwarf are totally removed. This is  due to them being tidally 
stripped over the course of several passages past the central region 
of NGC~1399. The nucleus, however, manages to survive. We also find that
the size and luminosity of the remnant are similar to those observed
for ultra-compact dwarf galaxies, if the simulated precursor nucleated
dwarf has a mass of $\sim$ $10^8$ $M_{\odot}$. These results suggest that
ultra-compact dwarf galaxies could have previously been more luminous 
dwarf spheroidal or elliptical galaxies with rather compact nuclei.
\end{abstract}


\section{Introduction}

Strong constraints on theoretical models of galaxy formation and evolution
have been provided by observational studies of the physical properties of
low-luminous and low surface-brightness dwarf spheroidal and irregular
galaxies in the field and in clusters (e.g., Ferguson \& Binggeli
1994; Mateo 1998). In detail, these studies have addressed such
observables as the scaling relation (Kormendy 1977;  Ferguson \& Binggeli
1994), the luminosity function (Binggeli, Sandage, \& Tammann
1985; Sandage, Binggeli, \& Tammann 1985), the presence of 
nuclear structures (Binggeli \& Cameron 1991),
and rotation-curve profiles (Moore 1994).
%
%
A new type of sub-luminous and extremely compact ``dwarf galaxy'' has
been recently discovered in an `all-object'  spectroscopic survey
centred on the Fornax Cluster (Drinkwater et al. 2000a, b).
These  have been already identified as bright compact objects (Hilker et al. 1999) 
and very luminous globular clusters around cD galaxies (Harris, Pritchet, \& McClure 1995).
These ``dwarf galaxies'', which are members of the Fornax Cluster,
have intrinsic sizes of $\sim$ 100 pc and absolute
$B$ band magnitude  ranging from $-13$ to $-11$ mag
and are thus called ``ultra-compact dwarf'' (UCD) galaxies.
The luminosities of UCDs are intermediate between those of globular
clusters and small dwarf galaxies and 
are similar to those of the bright end of the luminosity
function of the nuclei of nucleated dwarf ellipticals.
These UCDs are observed to be within 30 arcminutes of the central
dominant galaxy in Fornax, NGC~1399, and are distributed at larger radii
than this galaxy's globular cluster system. 

The purpose of this Letter is to suggest one possible origin for 
these newly discovered enigmatic UCDs. We adopt here the scenario that
UCDs are the stripped nuclei of dwarf galaxies 
and thereby investigate numerically how nucleated dwarf spheroidal galaxies
evolve dynamically under the strong tidal field of NGC~1399.
A growing number of evidences supporting this scenario have
been accumulating  for the case of $\omega$Cen and M54 (e.g., Majewski et al. 1999;
Layden \& Sarajedini 2000; van den Bergh 2000).   
We here demonstrate: (1)\,how a UCD is formed when a nucleated dwarf galaxy is
subjected to the strong tidal field of a massive galaxy such as NGC~1399, 
and (2)\,in what physical conditions this process of nucleated dwarf
galaxy $\rightarrow$ UCD formation can take place.
The importance of the tidal field of more massive galaxies 
in forming globular clusters (and even objects that are  an order of magnitude
brighter than globular clusters)  from nucleated dwarf galaxies has
already been discussed by several authors
(e.g., Zinnecker et al. 1988; Freeman 1993; Bassino et al. 1994).
Extraction of $only$ galactic nuclei from less massive galaxies 
by the tidal effects of more massive ones
is suggested to be important in a variety of different
contexts, such as the evolution of M32 and the formation of Galactic
halo globular clusters (Bekki et al. 2001).
We can think of and refer to this tidal effect as 
``galaxy threshing''.

\section{Model}

We consider a collisionless stellar system with a mass and 
size similar to that of nucleated dwarf galaxies, orbiting a massive 
elliptical galaxy (eg., NGC~1399) which is embedded within a massive dark
matter halo. To give our model a realistic radial density profile for
the NGC~1399 dark matter halo, we base it on both the X-ray observational
results of Jones et al. (1997) and the predictions from the standard cold
dark matter cosmogony (Navarro, Frenk, \& White 1996). The total mass of
NGC~1399 within 125 kpc (represented by M$_{\rm E}$) and the scale
length of the halo are  assumed to be $8.1 \times 10^{12} \rm M_{\odot}$
and 43.5 kpc, respectively.
By using two Plummer models (Binney \& Tremaine 1987) with rather different scale lengths,
we construct a model for nucleated dwarf galaxies as follows:
First, we place a smaller spherical stellar system having a
Plummer density profile with scale length $a_{\rm n}$
and mass M$_{\rm n}$ in the center of a larger spherical system
with a Plummer profile of scale length $a_{\rm d}$ and mass M$_{\rm d}$ 
($a_{\rm d}$ $\gg$ $a_{\rm n}$ and  M$_{\rm d}$ $\gg$ M$_{\rm n}$).
Second, in order to get the model to reach a new dynamical equilibrium,
we run the simulation of the dynamical evolution in the nested
Plummer models for 10 dynamical time scales.
Finally, we use the stellar system as the
model for a  nucleated dwarf galaxy.  
From now on, the outer (more massive) diffuse stellar component
and the inner (less massive) compact one are referred to as the
`envelope' and `nucleus', respectively.

The mass (luminosity) and the scale length of a dwarf is modeled according to
the observed scaling relation of Ferguson \& Binggeli (1994):
${\rm log}r_{0} [{\rm pc}] = -0.2 M_{\rm B} - 0.3 $ for 
bright dwarfs ($M_{\rm B}$ $<$ $-16$)
and ${\rm log}r_{0} [{\rm pc}] = -0.02 M_{\rm B} +2.6 $  
for faint ones ($M_{\rm B}$ $\ge$ $-16$), where
$r_{0}$ and $M_{\rm B}$ are the scale length  of the exponential
profile and  $B$ band absolute
magnitude, respectively. 
By assuming that $a_{\rm d}$ = $r_{\rm 0}$ and $M/L_{\rm B}$ (the ratio
of total stellar mass to total $B$ band luminosity) = 1.0 corresponding
to the observed value of the dwarf galaxy DDO 154 (Carignan \& Beaulieu 1989),
we determine $a_{\rm d}$ from the total mass ($=M_{\rm d} + M_{\rm n}$) of
the model.
Since the nuclei typically contributes about 2\% of
the total light of dwarfs (Binggeli \& Cameron 1991; Freeman 1993),
we assume that $M_{\rm n}/(M_{\rm d}+M_{\rm n})$ = 0.02 for all models.
Considering the fact that the central light excess with respect
to the adopted model profile is
observed to vary between nucleated dwarfs 
(e.g., Binggeli \& Cameron 1991; Ferguson \& Binggeli 1994),
we take the scale length ratio $a_{\rm n}/a_{\rm d}$ of the nested
Plummer models to be a free parameter.  
We investigate nucleated dwarf models in the luminosity range $-18$ $\le$
$M_{\rm B}$ $\le$ $-12$ and with $a_{\rm n}/a_{\rm d}$ ranging from 0.05
to 0.5. 

The orbit of a nucleated dwarf is assumed to be influenced
only by the gravitational potential resulting from the dark halo
component of NGC~1399. The center of NGC~1399 is always
set to be ($x$,$y$) = (0,0) whereas the initial position
and velocity of a dwarf are  ($x$,$y$) = ($r_{\rm in}$,0) 
and ($V_{\rm x}$,$V_{\rm y}$) = (0,$V_{\rm in}$),
respectively. By changing these two parameters
$r_{\rm in}$ and $V_{\rm in}$, we investigate how the
transformation process from dwarfs into  UCDs depends  on their orbits.  
Although we have investigated models with a variety of different 
$M_{\rm d}$, $a_{\rm d}$, $r_{\rm in}$, and  $V_{\rm in}$ values, 
we mainly describe here the results of a `standard' model
with total mass of M$_{\rm d}$ + M$_{\rm n}$ 
= 2.0 $\times$ $10^8$ M$_{\odot}$,
$a_{\rm d}$ = 7.94 $\times$ $10^2$ pc,  
$a_{\rm n}$ = 79.4 pc,
$r_{\rm in}$ = 200 kpc,
and $V_{\rm in}$ =  1.29 $\times$ $10^2$ km $\rm s^{-1}$.
Figure 1 shows the orbit  with respect to NGC~1399
and  the final mass distribution for 
the simulated dwarf galaxy  in the standard model. 

In the following, our units of mass, length, and time are
2.0 $\times$ $10^8$ M$_{\odot}$ (corresponding to M$_{\rm d}$ + M$_{\rm
n}$ in the standard model), 7.94 $\times$ $10^2$ pc ($a_{\rm d}$), 
and 2.36 $\times$ $10^7$ yr (dynamical time scale), respectively. 
Parameter values  and final morphologies for each model
are summarized in Table 1. The sixth and the seventh  columns give
orbital eccentricity ($e$) and  
pericenter distance ($r_{\rm p}$), 
respectively, for each model.
The eighth column 
describes the final morphological properties  after 160 time units 
(corresponding to 3.8 Gyr): 
`UCD' indicates a remnant with the envelope completely stripped yet 
the nucleus largely unaffected, `dE,N' the case where both the 
envelope and nucleus survive, and `no remnant' where both components
are tidally stripped. All the simulations have been carried out on a
GRAPE board (Sugimoto et al. 1990) in which 
energy and angular momentum are  conserved within 1 \% accuracy.
The total number of particles used for each model and the gravitational
softening length adopted for its envelope(nucleus) were 
10,000 (5000) and  0.47 (0.06) in our units, respectively.

\placefigure{fig-1}
\placefigure{fig-2}

\section{Results}

As the dwarf approaches the pericenter of its orbit for the first time 
($T$ = 20), the strong tidal field of NGC~1399 stretches the envelope
along the direction of the dwarf's orbit and consequently tidally strips
the stars of the envelope (see Figure 2).
Since the envelope loses a significant fraction of its mass after the
first passage of the pericenter, 
the envelope becomes more susceptible to the tidal effects of NGC~1399. 
As the dwarf again approaches the pericenter ($T$ $\sim$ 50 and 90), 
the envelope again loses a large number of its stars  owing to tidal stripping 
and consequently becomes less massive and more diffuse.
After four passages of the pericenter, the dwarf loses its envelope
almost entirely. 
The central nucleus, on the other hand, is just weakly influenced by the 
tidal force owing to its compact configuration during the tidal destruction
of the envelope. 
Thanks to its strongly self-gravitating nature,
the nucleus loses  only a small amount ($\sim$ 18 \%) of mass and thus
can keep its compact morphology during its tidal interaction with
NGC~1399. 

Figure 3 shows that a significant fraction
of the envelope of the dwarf is tidally stripped
every time it passes through the inner part of NGC~1399 (e.g., 63 \% between
$T$ = 25 and 60).
As a result of this, nearly all (98\%) of the stars within the dwarf's
envelope [radius $<a_{\rm d}$ (=0.8\,kpc)] are removed within four
passages of its pericenter
(corresponding to 3.8 $\times$ $10^9$ yr). 
The temporal increase of the envelope mass within $a_{\rm d}$ 
around  the apocenter is due to
the fact that a significant fraction of stripped
stars pass through the surrounds of the dwarf.
The ratio of the nuclear mass to the total mass is dramatically changed
from 0.08 (0.85) to 0.83 (1.0)  for $R$ $\le$ $a_{\rm d}$ ($R$ $\le$ 0.1$a_{\rm d}$), 
which implies that the final
remnant after ``threshing'' is nearly fully self-gravitating.
Thus a UCD of size $\sim$ 100 pc and mass $4.0\times 10^6$ M$_{\odot}$
(corresponding to $M_{\rm B}$ = $-12.4$ mag for M$/L_{\rm B}$ = 1.0)
is formed from the tidal interaction between a nucleated dwarf and
NGC~1399. It is clear from Figures 2 and 3 that 
before the formation of a UCD is completed, a compact nucleus with
a considerably diffuse outer envelope is seen (e.g., $T$ = 100). 
This suggests the existence of extremely diffuse nucleated dwarf
galaxies, formed in this intermediate stage of the dE,N$\rightarrow$UCD
conversion process. As is shown in Figure 4, the remnant shows a rather
compact density distribution because of it being composed mostly of
the self-gravitating nucleus at $T$ = 160.

Four important parameter dependences in the formation of UCDs
were found as follows (see the Table 1): For the model in which  
the nucleus of the dwarf is not so compact ($a_{\rm n}/a_{\rm d}$ $\sim$
0.5; Model 2), both the nuclear and envelope components disintegrate
under the influence of the tidal field and, accordingly, 
no remnant is left. A UCD is not formed in Model
3 with larger values of $r_{\rm in}$ (200 kpc) and $r_{\rm p}$ (95 kpc),
but a smaller ellipticity ($e =0.35$), whereas one is in Model 4 with a
smaller $r_{\rm in}$ (100\,kpc) and a smaller $e$ (0.34).
This suggests that nucleated dwarfs with either a smaller $r_{\rm p}$
(or smaller $e$) or smaller $r_{\rm in}$, 
are more likely to be transformed into a UCD and thus 
UCDs should show a centrally concentrated spatial distribution around
NGC~1399. Thirdly, the formation of a UCD seems to have no dependence
on the mass of the nucleated dwarf's envelope, M$_{\rm d}$, as shown
by Models 1, 5-7. Irrespective of the masses of these models with $e$ =
0.81 and $r_{\rm p}$ = 21 kpc, UCDs are formed by ``galaxy threshing''.
This indicates that UCDs observed to have different masses were
previously the nuclei of nucleated dwarfs with different masses.
Finally, for the model with a rather small mass for NGC~1399 
(Model 8), disintegration of the simulated nucleated dwarf
does not occur at all, even if $r_{\rm p}$ is small. This suggests that
UCDs are formed only in the surrounds of massive galaxies.

Several observational properties of nucleated dwarf galaxies
(e.g., Freeman 1993) and numerical simulations (Bassino et
al. 1994) suggest that globular clusters are the stripped nuclei of dwarf
galaxies. We have confirmed the ``disintegration scenario'' (Bassino et al. 1994)
in  which  UCDs ($-13$ $\le$  $M_{\rm B}$
$\le$ $-11$ mag), which are more than an order of magnitude brighter
than the Galactic globular clusters, were also previously nuclei of
more massive dwarf galaxies (M$_{\rm d}$ $\sim$ $10^8$
M$_{\odot}$) orbiting NGC~1399. Accordingly, it is likely that
physical properties of UCDs such as the luminosity-size relation and
stellar content are more like those of globular clusters than those of
dwarf spheroidal/elliptical galaxies: it might not be correct to call the
observed compact objects around NGC~1399 `` dwarf galaxies''.
Our numerical simulations with variously different parameters
revealed that not all of nucleated dwarfs can be transformed 
into UCDs by ``galaxy threshing'' (e.g., for the case of the smaller ratio
of the dwarf mass to  the giant mass),
which suggests that the number ratio of UCDs to nucleated dwarfs
is different between different environments.

\placefigure{fig-3}
\placefigure{fig-4}

\section{Conclusion}

The present study provides the following two implications as to the
nature of UCDs. First, the stellar populations of UCDs are unlikely to be
young even if the precursor nuclei of dwarf galaxies have obviously young
populations. This is essentially because more than a few Gyrs,
which is enough for young stars to become intermediate-age populations
due to aging,   are necessary for ``galaxy threshing''
to transform a nucleated dwarf into an UCD. 
Second, the UCD luminosity function is not necessarily similar
to that of the nuclei of nucleated dwarf galaxies, because ``galaxy
threshing'' is a selective process: extraction of the nucleus 
depends strongly on the orbits and masses of the dwarfs. 
Since these physical properties can be directly observed in future
observations -- not only in the Fornax cluster but also in other nearby
clusters such as Coma and Virgo -- they will provide a critical test
of the viability of the ``threshing'' scenario for UCD formation.

\acknowledgments

We are  grateful to the anonymous referee for valuable comments,
which contribute to improve the present paper.

\clearpage


\figcaption{
The orbital evolution  of the simulated dwarf with respect to
the center of NGC~1399 (left)
and final mass distribution of the dwarf at $T$ = 160 in our units  (right). 
The center of NGC~1399 is set to be always ($x$,$y$) = (0,0). 
Here the scale is given in our units (0.8 kpc) and thus
each frame measures 448 kpc. 
Filled circle represents the position of the dwarf 
at $T$ = 0 and  the orbital evolution is indicated by open circles with
the time interval of 20 time units corresponding to  
4.7 $\times$ $10^8$ yr ($T$ = 0, 20, 40,
60, 80, 100, 120, 140, and 160). 
Note that owing to the strong tidal field of NGC~1399,
the dwarf is greatly stretched and most of the outer stellar components 
of the dwarf is tidally stripped away from it at $T$ = 160 (3.8 Gyr). 
\label{fig-1}}

\figcaption{
Morphological evolution projected onto $x$-$y$ plane
for  the envelope (upper six panels) and the nucleus (lower)
in the simulated dwarf. The detailed
explanation for  the definition of the envelope
and the nucleus is given in the main manuscript.
The time indicated in the upper left
corner of each frame is given in our units (2.36 $\times$ $10^7$ yr) and
each frame measures 19.2 kpc for the upper six (envelope)
and 1.52 kpc for the lower (nucleus).
Note that nearly all of stars initially in the envelope
of the dwarf are tidally stripped away whereas
the nucleus keeps  its initial compact configuration.
This result clearly demonstrates that the strong tidal
field of a giant galaxy can transform a nucleated dwarf
into a very compact
galaxy.
\label{fig-2}}

\figcaption{
$Upper$: Time evolution of the total mass within $R$ $<$ 1 (=$a_{\rm d}$) in our units,
where $R$ is the distance from the center of the dwarf, for the envelope
(solid) and the nucleus (dotted).
Here the mass normalized by the initial mass ($T$ = 0) within $R$ $<$ 1
is plotted both for the envelope and the nucleus.
Therefore the solid (dotted) line  describes what  fraction of stars initially
in the envelope (nucleus) is  removed from the system at each time.
Note that the envelope's stars are preferentially stripped
during tidal interaction between NGC~1399 and the dwarf.
Note also that although only $\sim$ 7.5 \% of stars in the nucleus
is tidally stripped at $T$ = 160, nearly all ($\sim$ 98 \%) of stars
in the envelope are stripped.
$Lower$: Time evolution of the ratio of nucleus mass ($M_{\rm n}$) to total one 
($M_{\rm d}+M_{\rm n}$) 
for $R$ $<$ 1.0 (solid) and $R$ $<$ 0.1 (dotted).
This figure describes how strongly the nucleus becomes self-gravitating
at each time.
It is clear that as the envelope is gradually removed,
the nucleus becomes more strongly self-gravitating
not only in the inner part ($R$ $<$ 0.1) but also in the outer one ($R$ $<$ 1). 
\label{fig-3}}

\figcaption{
Density profiles  of the simulated dwarf at $T$ = 0 (left) and 160 (right). 
The profiles for all, the envelope, and the nucleus
components are given by solid, dotted, and dashed lines, respectively.
Note that 
the  nucleus component dominates almost exclusively  
the final density profile of the simulated dwarf.
\label{fig-4}}

\clearpage

\begin{deluxetable}{cccccccc}
\footnotesize
\tablecaption{Results of different models of tidal interaction between
a dE,N and a giant galaxy \label{tbl-1}}
\tablewidth{0pt}
\tablehead{
\colhead{model number} 
& \colhead{$M_{\rm E}$ ($10^{13} M_{\odot}$)} 
& \colhead{$M_{\rm d}$ ($10^8 M_{\odot}$)} 
& \colhead{$a_{\rm n}/a_{\rm d}$} 
& \colhead{$r_{\rm in}$ (kpc)} 
& \colhead{$e$} 
& \colhead{$r_{\rm p}$ (kpc)} 
& \colhead{final morphology}}
\startdata
1 & 8.1 & 2.0 & 0.1 & 200 & 0.81 & 21 & UCD \\
2 & 8.1 & 2.0 & 0.5 & 200 & 0.81 & 21 & no remnant \\
3 & 8.1 & 2.0 & 0.1 & 200 & 0.35 & 95 & dE,N \\
4 & 8.1 & 2.0 & 0.1 & 80 & 0.34 & 39 & UCD \\
5 & 8.1 & 0.05 & 0.1 & 200 & 0.81 & 21 & UCD \\
6 & 8.1 & 0.31 & 0.1 & 200 & 0.81 & 21 & UCD \\
7 & 8.1 & 12.2 & 0.1 & 200 & 0.81 & 21 & UCD \\
8 & 0.8 & 2.0 & 0.1 & 200 & 0.81 & 21 & dE,N \\
\enddata

\end{deluxetable}


\begin{thebibliography}{}


\bibitem[Bassino et al. 1994]{bas94}
Bassino, L. P., Muzzio, J. C., \& Rabolli, M. 1994, \apj, 431, 634


\bibitem[Bekki et al. 2001]{be01}
Bekki, K., Couch, W. J., \& Drinkwater, M. J. 2001, in preparation


\bibitem[Binggeli et al. 1985]{bi85}
Binggeli, B., Sandage, A., \& Tammann, G. A. 1985, \aj, 90, 1681 

\bibitem[Binggeli \& Cameron 1991]{bi91}
Binggeli, B., \& Cameron, L. M., 1991, \aap, 252, 27

\bibitem[Binney \& Tremaine 1987]{bi87}
Binney, J., \& Tremaine, S., 1987 in Galactic Dynamics.



\bibitem[Carignan \& Beaulieu 1989]{bi89}
Carignan, C.,  \& Beaulieu, S.  1989, \apj, 347, 760



\bibitem[Drinkwater  et al.  2000a]{dr00a}
Drinkwater, M. J., Phillipps, S.,  Jones, J. B., Gregg, M. D.,   Deady, J. H., 
Davies, J. I., Parker, Q. A., Sadler, E. M., \& Smith, R. M. 2000a,
\aap, 355, 900 

\bibitem[Drinkwater  et al.  2000a]{dr00b}
Drinkwater, M. J., Jones, J. B., Gregg, M. D., \& Phillipps, S.  2000b,
PASA, in press



\bibitem[Ferguson \& Binggeli  1994]{fb94}
Ferguson, H. C., \& Bingelli, B. 1994, A\&ARv, 6, 67


\bibitem[Freeman 1993]{fr93}
Freeman, k. C. 1993, in The globular clusters-galaxy connection,
edited by Graeme H. Smith, and Jean P. Brodie,
ASP conf. ser. 48, p608 



\bibitem[Harris et  al. 1995]{ha95}
Harris, W. E., Pritchet, C. J.,   \& McClure, R. D.,
1995, \apj, 441, 120 

\bibitem[Hilker et al.  1999]{hi95}
Hilker, M., Infante, L., \&  Richtler, T. 1999, \aaps, 138, 55


\bibitem[Jones et al 1997]{jo97}
Jones, C., Stern, C., Forman, W., Breen, J., David, L., Tucker, W., \&
Franx, M. 1997, \apj, 482, 143


\bibitem[Kormendy  1977]{ko77}
Kormendy, J. 1977, \apj, 218, 333

\bibitem[Layden \& Sarajedini 2000]{ls00}
Layden, A. C., \&  Sarajedini, A. 2000, \aj, 119, 1760

\bibitem[Majewski et al.  2000]{ma00}
Majewski, S. R.  et al. 1999,  
The  Galactic Halo : From Globular Cluster to Field Stars, 
Proceedings of the 35th Liege International Astrophysics Colloquium, 
Edited by A. Noels, P. Magain, D. Caro, E. Jehin, G. Parmentier, and A. A. Thoul. 
p619

\bibitem[Mateo 1998]{ma98}
Mateo, M. 1998, \araa, 36, 435

\bibitem[Moore  1994]{mo94}
Moore, B. 1994, \nat, 370, 629



\bibitem[Navarro et al. 1996]{na96}
Navarro, J. F., Frenk, C. S., \& White, S. D. M.
1996, \apj, 462, 563








\bibitem[Sandage et al. 1985]{sa85}
Sandage, A., Binggeli, B., \& Tammann, G. A. 1985, \aj, 90, 1759

\bibitem[Sugimoto et al. 1990]{sug90}
Sugimoto,~D., Chikada,~Y., Makino,~J., Ito,~T., Ebisuzaki,~T., \&
Umemura, M. 1990, \nat, 345, 33

\bibitem[van den Bergh 2000]{va00}
van den Bergh, S. 2000, \apj, 530, 777 

\bibitem[Zinnecker  et al. 1988]{zi88}
Zinnecker, H., Keable, C. J., Dunlop, J. S., Cannon, R. D., \&  Griffiths,  W. K.
1988, in Grindlay, J. E., Davis Philip A. G., eds, Globular cluster systems in Galaxies,
Dordrecht, Kluwer, p603





\end{thebibliography}
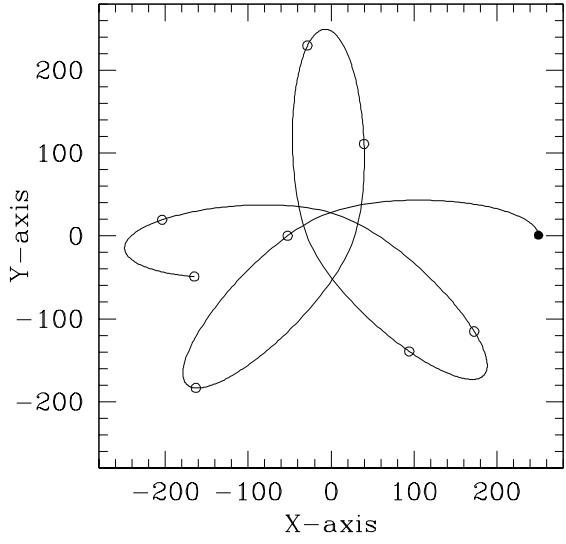
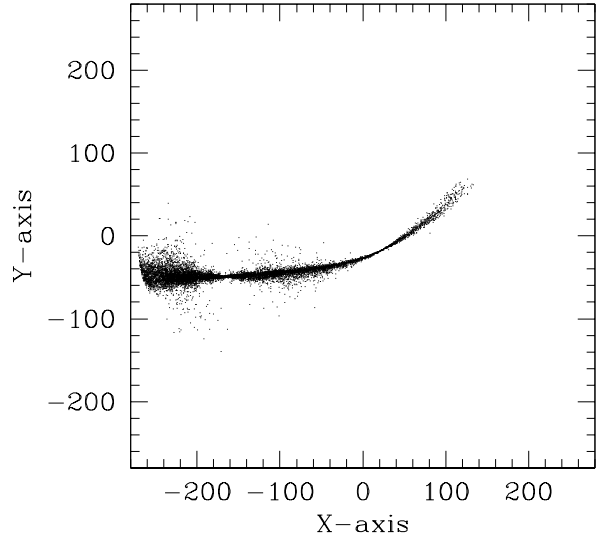
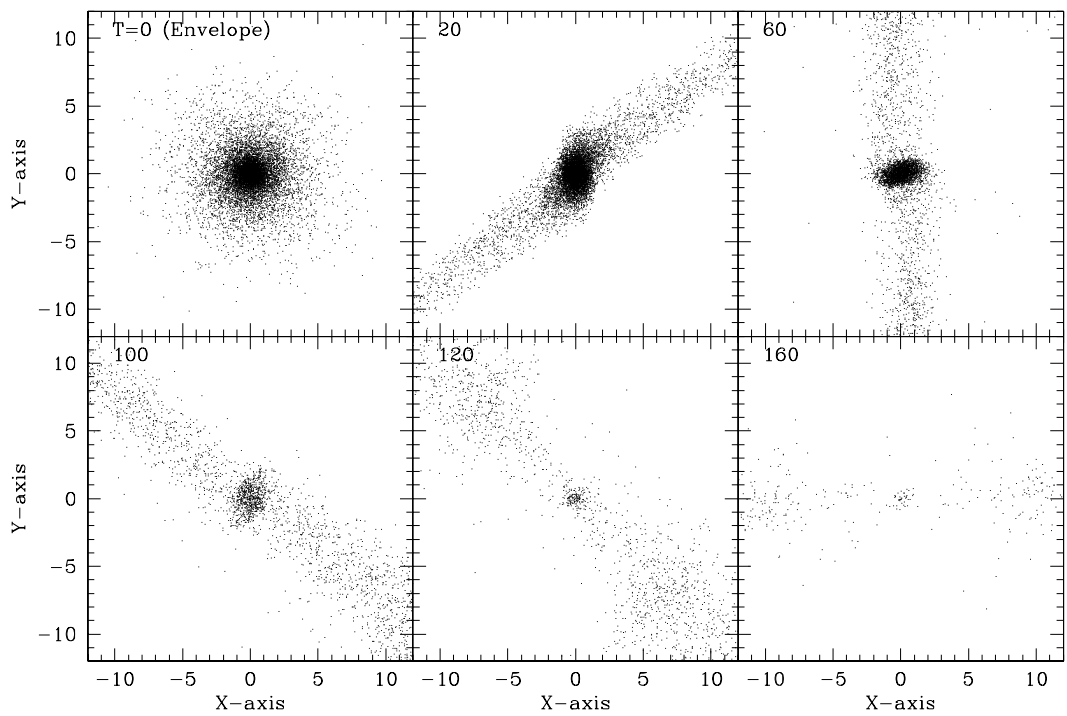
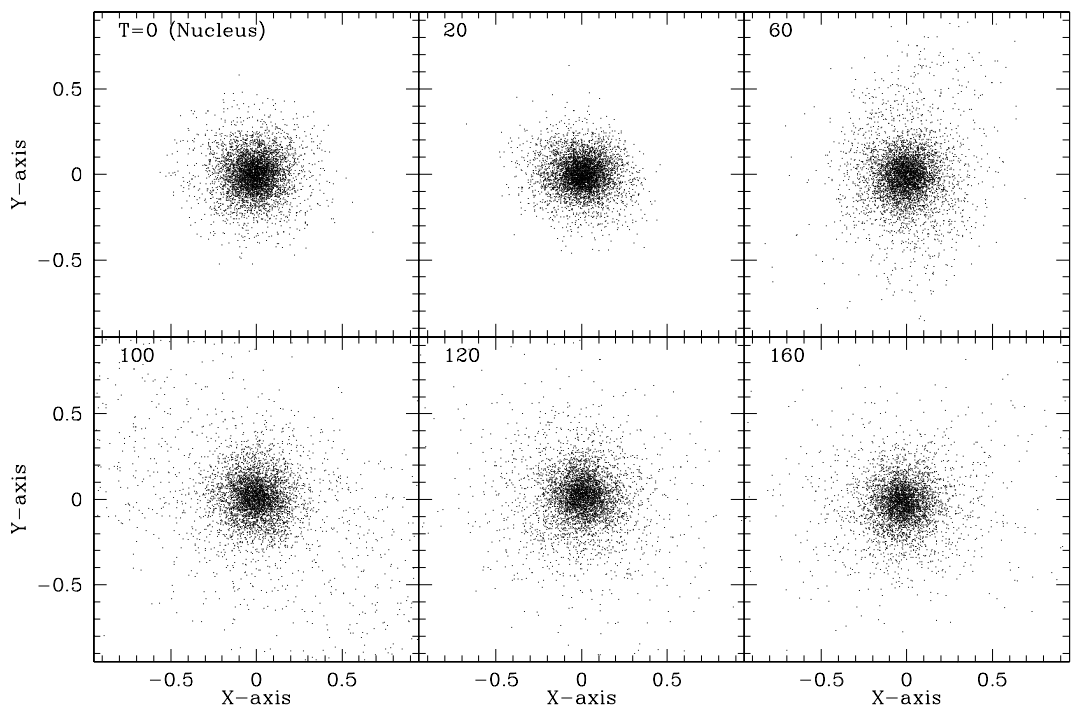
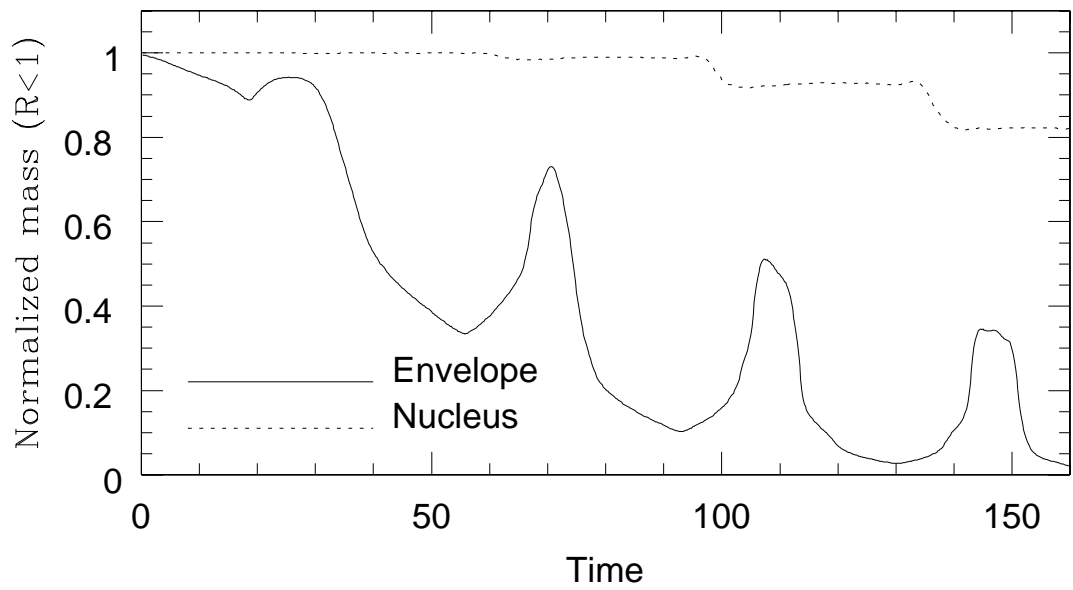
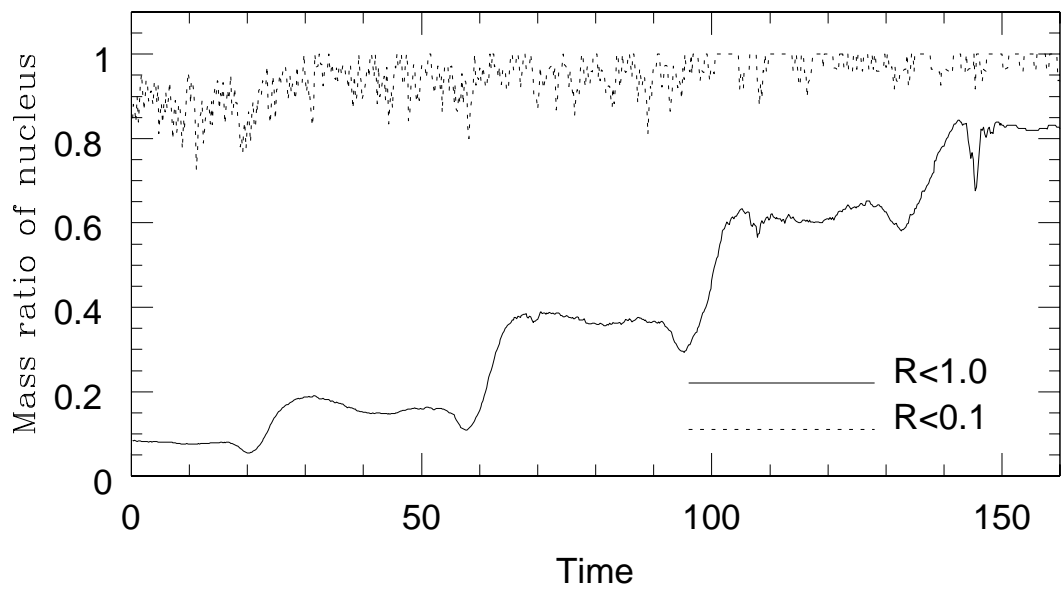
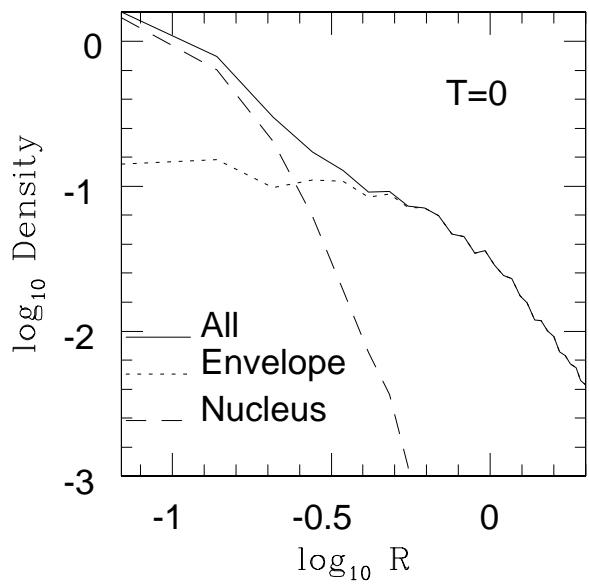
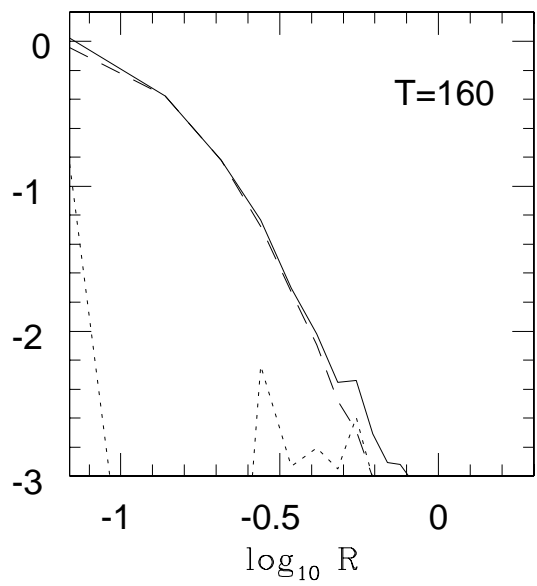
\end{document}